\documentstyle[aps,prb,multicol,graphicx]{revtex}

\begin{document}

\draft

\title{Thermal Conductivity of Mg-doped CuGeO$_3$}

\author{J. Takeya, I. Tsukada and Yoichi Ando}
\address{Central Research Institute of Electric Power
Industry, Komae, Tokyo 201-8511, Japan}

\author{T. Masuda
and K. Uchinokura
}
\address{Department of Advanced Materials Science,
The University of Tokyo,
7-3-1, Hongo, Bunkyo-ku, Tokyo 113-8656, Japan}

\date{Received }

\maketitle

\begin{abstract}
The thermal conductivity $\kappa$ is measured
in a series of Cu$_{1-x}$Mg$_x$GeO$_3$ single crystals
in magnetic fields up to 16~T.
It has turned out that heat transport by spin excitations
is coherent for lightly doped samples,
in which
the spin-Peierls (SP) transition exists,
at temperatures well above the SP transition temperature
$T_{\rm SP}$.
Depression of this spin heat transport appears
below $T^*$, at which the spin gap {\it locally} opens.
$T^*$ is not modified with Mg-doping
and that $T^*$ for each Mg-doped sample remains
as high as $T_{\rm SP}$ of pure CuGeO$_3$,
in contrast to $T_{\rm SP}$ which is strongly suppressed
with Mg-doping.
The spin-gap opening
enhances phonon part of the heat transport
because of reduced scattering by the spin excitations,
producing an unusual peak.
This peak diminishes when the spin gap is suppressed
both in magnetic fields and with the Mg-doping.

\end{abstract}


%
\begin{multicols}{2}
\narrowtext

\section{Introduction}

The discovery of the first inorganic spin-Peierls (SP) compound
CuGeO$_3$ (Ref.~\onlinecite{hase1})
has triggered subsequent studies of the
impurity-substitution effect on the spin-singlet states,
\cite{hase2} and the existence of the disorder-induced transition
into three-dimensional antiferromagnetic (3D-AF) state has been
established. \cite{hase3,oseroff}
Since the relevant
exchange energies, i.e., intra- and interchain coupling $J$ and
$J'$, do not change with doping except at the impurity sites,
it is difficult to understand the impurity-induced transition from
the SP to 3D-AF state
in the framework of the \lq\lq conventional"
competition between dimensionalities
(where $J'/J$ is an essential parameter \cite{inagaki})
and therefore such impurity-substitution effect is a matter of
current interest.

In CuGeO$_3$, a small amount of impurity leads
to an exotic low-temperature phase where the lattice dimerization
and antiferromagnetic staggered moments simultaneously appear
[dimerized AF (D-AF) phase]. \cite{masuda}
Moreover, when the
impurity concentration $x$ exceeds a critical concentration $x_c$, the
SP transition measured by dc
susceptibility disappears \cite{sasago,martin} and a uniform AF
(U-AF) phase appears below the N\'{e}el temperature $T_N$ $\sim$ 4~K.
\cite{masuda}
The D-AF ground state can be understood as
a state of spatially modulated staggered moments accompanied
with the lattice distortion. \cite{fukuyama1,kojima}
However,
the mechanism of
the depression of the SP phase and the establishment of
the disorder-induced antiferromagnetism
are still to be elucitated.
A transport measurement could be a desirable tool
in dealing with such a problem,
because the mobility of spin excitations or
spin diffusivity
is often sensitive to the impurities.
Nevertheless, even a crude estimation of the mobility
has not been carried out for the spin excitations so far.

Recently, we reported intriguing behaviors of the thermal
conductivity $\kappa$ of pure CuGeO$_3$. \cite{ando}
In Ref.~\onlinecite{ando}, the existence of
the spin heat channel ($\kappa_s$)
was suggested.
However, we cannot separate $\kappa_s$ and the phonon
thermal conductivity $\kappa_{\rm ph}$
with the measurement of the pure sample only.
In this work, we have resolved
this problem with Mg-doped crystals.
It turned out that the spin diffusion length,
which can be calculated from $\kappa_s$,
is much larger than the distance
between adjacent spins for the pure CuGeO$_3$,
indicating coherent heat transport
due to the spin excitations.
Since the spin heat transport
almost disappears for the heavily doped samples
in which the long-range SP ordering is absent,
it is suggested that the large mobility of the spin excitations
plays an important role in the SP ordering.

The spin heat transport is also a probe of the spin gap.
It is found that local spin-gap opening
is robust against the impurity doping,
though the temperature of the long-range ordering
is strongly suppressed with $x$.
Phonon heat channel ($\kappa_{\rm ph}$) provides an
unusual thermal conductivity peak in the SP state,
which was one of the topics of our previous report.\cite{ando}
The peak is drastically suppressed with
both magnetic field and Mg-doping,
indicating that the spin-gap opening has produced the peak in
$\kappa_{\rm ph}$
via strong phonon-spin interaction.

\section{Experimental}

The single crystals of Cu$_{1-x}$Mg$_x$GeO$_3$ were
grown with a floating-zone method. The Mg concentration is
determined by inductively coupled plasma-atomic emission
spectroscopy (ICP-AES). \cite{masuda}
The critical concentration $x_c$ was carefully determined
by
Masuda {\it et al.} for the same series of crystals that we have used.
\cite{masuda,masuda2}
They found that
the N\'{e}el temperature $T_N$ jumps at the impurity-driven
transition from D-AF to U-AF phase
and that phase is separated
in the transition range of $0.023 \leq x \leq 0.027$,
indicating that the transition is of first order.\cite{masuda,masuda2}
We used six samples with $x = 0$, $0 < x < x_c$, and $x > x_c$
for the thermal conductivity measurement.
They were originally prepared for the neutron and
the synchrotron x-ray experiments,
and thus the sample homogeneity has been confirmed in detail
which is described elsewhere.\cite{masuda2}
Typical sample size is 0.2 $\times$ 2 $\times$ 7 mm$^3$
($a \times b \times c$).
Since our interest is thermal conductivity along the 1D spin chain,
we need the sample with a longer dimension along the $c$ axis.
Thus we payed particular attention to the Mg homogeneity
along the $c$ axis for this study,
and the best way is to use the samples prepared for the
neutron experiments.
The fundamental magnetic properties of the samples used already
appeared in another paper.\cite{masuda2}

The thermal conductivity is
measured using a \lq\lq one heater, two thermometers" method as
the previous measurement.\cite{ando}
The base
of the sample is anchored to a copper block held at desired
temperatures. A strain-gauge heater is used to heat the sample.
A matched pair of microchip Cernox thermometers are carefully
calibrated and then mounted on the sample.
The maximum temperature
difference between the sample and the block is 0.2~K.
Typical
temperature difference between the two thermometers is $\sim$0.5~K
above 20~K and 0.2~K below 20~K.
The temperature-controlled block is covered with
a metal shield
so that heat loss through radiation becomes negligible.
Magnetic field is applied parallel to the $c$-axis direction.
The measured temperature range is below 30~K,
which is well below the temperature scale of
the intra-chain coupling
$J_c$ ($\sim$~120~K according to neutron scattering measurement
\cite{nishi}),
and therefore,
the one-dimensional quantum spin liquid (1D-QSL) is
realized above the transition temperatures.

\section{Results}

Figure~1 shows the temperature dependence of $\kappa$
for samples with and without Mg substitution:
1(a)-1(f) are for $x = 0,\ 0.016$,
0.0216, 0.0244, 0.032 and 0.040, respectively.
$x = 0.016$ is well below $x_c$,
$x = 0.0216$ is close to $x_c$, $x = 0.0244$ is on the
border and $x = 0.032$ and $x = 0.040$ are in the U-AF region.
The transition temperatures $T_{\rm SP}^{\chi}$ for the
$x \leq x_c$ samples and $T_N$ for the $x > x_c$ samples,
which were determined by susceptibility measurement, \cite{masuda2}
are indicated by arrows in the figures.
A sharp drop is observed just below $T_{\rm SP}^{\chi}$
for the pure CuGeO$_3$
as presented in the inset of Fig.~1(a).
The $x = 0.016$ sample shows the same behavior at $T_{\rm SP}$,
while $\kappa$ is {\it enhanced} just below $T_{\rm SP}$
for the $x = 0.0216$ and 0.0244 samples.

\begin{figure}
\includegraphics[width=8cm]{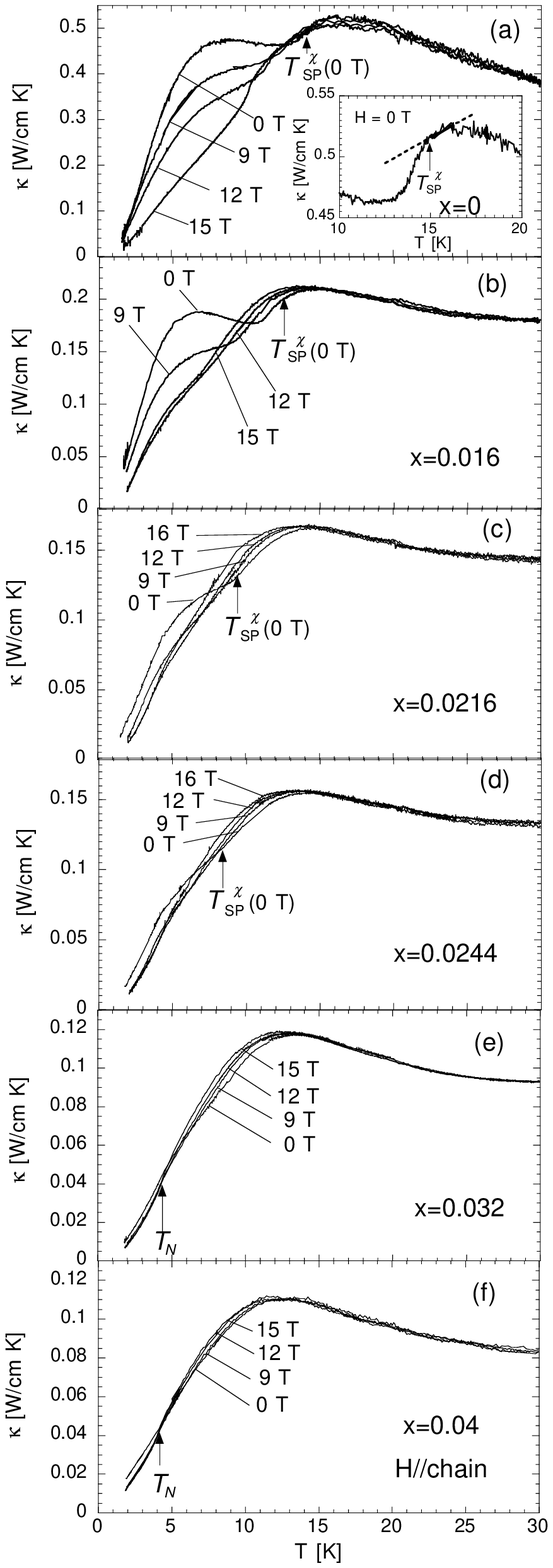}
\caption{Thermal conductivity of Cu$_{1-x}$Mg$_x$GeO$_3$ single
crystals in magnetic fields up to 15 or 16 T applied along the
chain direction, for (a) $x = 0$, (b) $x = 0.016$, (c) $x =
0.0216$, (d) $x = 0.0244$, (e) $x = 0.032$, and (f) $x = 0.040$.
Inset: a magnified view of $\kappa$ near the spin-Peierls
transition in zero field. The dashed line is an extrapolation of
$\kappa$ from above the transition temperature } \label{fig1}
\end{figure}

In Figs.~1(e) and 1(f), a slight change in the curvature is
observed at $T_N$ and the field dependence in the N\'{e}el state
is opposite to that above $T_N$; $\kappa$ decreases with field
below $T_N$ while $\kappa$ increases with field above $T_N$. The
three-dimensional magnon excitations, which appear in the N\'{e}el
state, are responsible for this behavior. The thermal conductivity
in U-AF phase will be further discussed elsewhere.\cite{takeya3}
Although the N\'{e}el transition (to D-AF phase) should be present
also for the $x \leq x_c$ samples, we cannot identify the
corresponding feature at the transition, because $T_N$ ($\sim
2-2.5$~K) \cite{masuda} is too close to the lower temperature
limit of our measurement, $\sim$~2~K.

Taking a look at the full scale of each figure,
one can notice that
$\kappa$ is strongly suppressed with Mg substitution
in the whole temperature range.
Two thermal conductivity peaks are present in the zero-field curve
of Fig.~1(a);
one is in the SP phase and the other is in the 1D-QSL
state.
The broad shape of the high-temperature peak is maintained
with doping.
On the other hand,
the low-temperature peak (SP peak) diminishes
with $x \le x_c$, and disappears for the
$x > x_c$ samples.

The SP peak is suppressed also in magnetic fields
[Figs.~1(a)-1(d)].
In order to see the field dependence in more detail,
the thermal conductivity at the SP peak is
plotted as a function of magnetic field in Fig.~2
for the $x = 0$ sample.
$\kappa$ is suppressed
with magnetic field and suddenly drops at the threshold field, $H_c$,
at which the system makes a transition to the incommensurate phase.
The same features were observed
in our previous
measurement with the crystal grown by
the laser-heated pedestal growth technique.\cite{ando}
Also, the typical value of $\kappa$, for example
at the high-temperature peak, is
almost the same as that of the previous result.
We thus believe that
the above features are intrinsic in this material
regardless of the growth method of the crystal,
as long as high quality single crystal is used.\cite{vasilev}

\begin{figure}
\includegraphics[width=0.8\columnwidth]{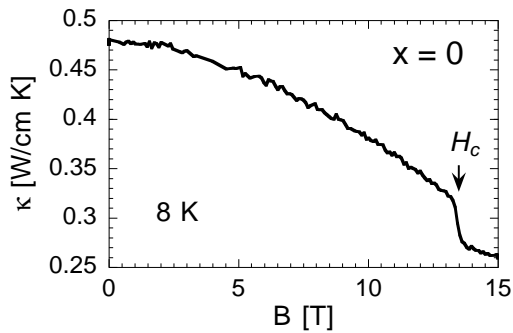}
\caption{Magnetic field dependence of the thermal conductivity at
8~K for pure CuGeO$_3$, demonstrating the suppression of the
low-temperature peak with field application. } \label{fig2}
\end{figure}

In a temperature range close to $T_{\rm SP}^{\chi}$, $\kappa$ {\it
increases} with field for the Mg-doped samples, while little field
dependence is seen in the pure sample [Figs.~1(a)-1(d)]. Note that
the field dependence extends even above $T_{\rm SP}^{\chi}$ for
the Mg-doped samples. The variation in the field dependence
implies qualitative difference of the SP transition between pure
CuGeO$_3$ and doped CuGeO$_3$. This field dependence is apparent
also in the $x = 0.032$ sample, which is above $x_c$ [Fig.~1(e)],
but rapidly diminishes in the heavily doped sample [Fig.~1(f)].

\section{Discussion}

CuGeO$_3$ is an insulator and heat conduction by
electrons or holes is absent unlike metals.
Instead, low-energy spin excitations\cite{cloizeaux}
can carry heat.
In the 1D-QSL state,
total $\kappa$ is given by a sum of $\kappa_s$ and
$\kappa_{\rm ph}$ as,
\begin{equation}
\kappa = \kappa_s + \kappa_{\rm ph}
\ \ \ ({\rm in\ 1D-QSL},\ T_{\rm SP} < T \ll J).
\end{equation}
In contrast, $\kappa_s$ is suppressed
at the lowest temperature region
in SP phase because
few spin excitations are present due to the spin gap,
and the total $\kappa$ represents a phonon contribution only,
\begin{equation}
\kappa \sim \kappa_{\rm ph}\ \ \  (T \ll T_{\rm SP}).
\end{equation}
The field and impurity dependence of $\kappa_{\rm ph}$
below $T_{\rm SP}$ will be discussed in subsection A.
The value of $\kappa_s$ in 1D-QSL will be crudely estimated
in subsection B.
Finally, the problem of local spin-gap formation
will be dealt with in subsection C,
using this $\kappa_s$ as a probe.

\subsection{$\kappa$ below $T_{\rm SP}$:
depression of the SP peak in magnetic fields and with Mg-doping}

Since the low-temperature peak is drastically suppressed
with the application of field as shown in Figs.~1 and 2,
the feature should be related to the spin-excitation
spectrum.
We have proposed the following explanation for this peak.\cite{ando}
At temperatures well below $T_{\rm SP}$,
where low-energy spin excitations are negligible,
heat is carried mostly by phonons
and
$\kappa_{\rm ph}$ increases with temperature
because of the growing population of phonons
like usual insulating crystals.
With increasing temperature, the
increasing number of thermally excited spin excitations
scatter phonons, and $\kappa_{\rm ph}$ diminishes.
As a result,
the thermal conductivity peak shows a peak.
Although the thermal spin excitations begin to carry heat,
the increase in $\kappa_s$ appears only in the temperature region
slightly below
$T_{\rm SP}$ [see inset in Fig.~1(a)].

In order to confirm the above understanding,
the impurity substitution is helpful
because the spin gap is absent in heavily substituted samples.
We normalized the size of the peak for each sample
as $\kappa$(5~K)/$\kappa$(15~K),
and
plotted it against $x$ in Fig.~3.
One can see the suppression of the peak with $x$
similarly to the case of field application (Fig.~2).
Moreover, the peak disappears at the concentration
$x > x_c$ where the long-range SP order no longer exists,
as shown in Figs.~1(e) and 1(f).
We have thus confirmed the correlation
between spin gap and the peak also by Mg substitution,
giving the additional evidence for
the scenario described in the previous paragraph.

\begin{figure}
\includegraphics[width=0.8\columnwidth]{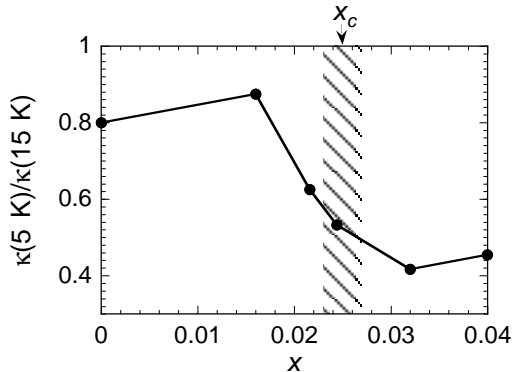}
\caption{A normalized size of the low-temperature peak,
$\kappa$(5~K)/$\kappa$(15~K), is plotted against Mg-concentration.
A suppression of the peak with Mg doping is evident. }
\label{fig3}
\end{figure}

The $x = 0.0244$ sample, which is on the border from SP (D-AF) to
U-AF demonstrates enhancement of $\kappa$ below $T_{\rm SP}^\chi$,
as observed in Fig.~1(d). Since the enhancement is suppressed in
magnetic fields, the existence of a well-defined spin gap is
suggested even at $x = x_c$. We can notice the rapid change in
$\kappa_s$ in the (low-temperature) vicinity of $T_{\rm SP}$,
which is another sign of the spin-gap opening, for the $x = 0$ and
0.016 samples [Figs.~1(a) and 1(b)]. However, this feature is
absent for the $x = 0.0216$ and 0.0244 samples [Figs.~1(c) and
1(d)], since $\kappa_s$ is significantly reduced with $x$ because
of impurity scattering.

Figure~4(a) shows the zero-field thermal conductivity of
all the samples.
The thermal conductivity is strongly suppressed
with Mg-substitution in the whole measured temperature
range.
The $x$ dependence of $\kappa$ below $\sim$~4~K
can be understood as the difference in the scattering rate of phonons.
Since
we showed in Ref.~\onlinecite{ando}
that the scattering by planer defects is dominant
for the $x = 0$ sample,
difference in the number of planer defects
may cause
the variation of $\kappa_{\rm ph}$ in our series of samples.
The number of planer defects does not necessarily
increase with increasing $x$;
note that $\kappa$ in the $x = 0.032$ sample is smaller
than that in the $x = 0.040$ sample below $\sim$~5~K.

\begin{figure}
\includegraphics[width=0.8\columnwidth]{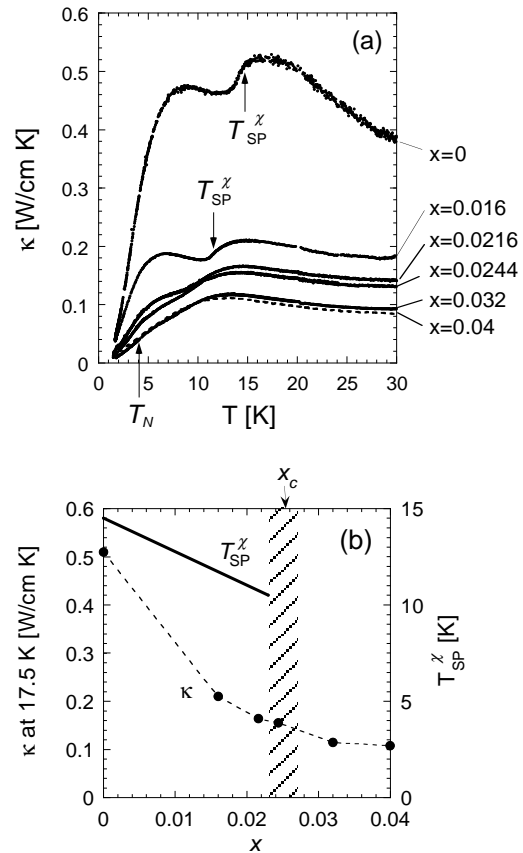}
\caption{(a) Temperature dependence of the zero-field thermal
conductivity of Cu$_{1-x}$Mg$_x$GeO$_3$ single crystals. (b) The
thermal conductivity at 17.5 K as a function of Mg concentration.
The spin-Peierls temperatures measured by dc susceptibility are
plotted together.} \label{fig4}
\end{figure}

\subsection{$\kappa$ above $T_{\rm SP}$:
coherent heat transport in the 1D-QSL}

Remembering the origin of the low-$T$ thermal conductivity peak,
we can notice that scattering by the spin excitations,
in turn, becomes dominant in $\kappa_{\rm ph}$
above the peak temperature
for the $x \leq x_c$ samples.
Noting that
all the $x \geq 0.016$ curves looks parallel to one another
above $\sim$~15~K,
$\kappa_{\rm ph}$ in the $x > x_c$ samples should probably
have a temperature dependence similar to that in the
$x = 0.016$, 0.0216 and 0.0244 samples above $\sim$~15~K.
Therefore, it is expected that phonon heat transport is
governed by the spin-scattering also for the $x > x_c$ samples,
at least above $\sim$~15~K.

In contrast to rather complicated temperature dependence
below $T_{\rm SP}$,
$\kappa(T)$ in the 1D-QSL region is simpler,
which is an advantage of discussing $x$ dependence
in the region.
As a crude approximation, it is assumed that
the phonon part is not so much $x$ dependent
and that most of the $x$-dependence comes from $\kappa_s$,
because the detailed characterization \cite{masuda2}
guarantees good homogeneity
in both pure and Mg-doped crystals and
the direct modification in the phonon modes
due to the Mg substitution
can be estimated to be negligible.
The scattering rate of phonons due to the point defects
is more than two orders of magnitude smaller than the total
scattering rate.\cite{berman}
One may notice that
the robustness of $\kappa_{\rm ph}$ to the Mg-doping
is natural,
considering that
phonons are mainly scattered by spin excitations above $\sim$~15~K,
which is the conclusion of the previous paragraph.
Since the population of both phonons and spin excitations
does not change so much with $x$ in the temperature range
above $T_{\rm SP}$,
as has been reported in specific heat results,
\cite{oseroff,lorenz,masuda3}
$x$ dependence of $\kappa_{\rm ph}$ is expected to be small
in the 1D-QSL state.\cite{defect}

Figure~4(b) shows the $x$ dependence of $\kappa$ at 17.5 K.
$\kappa$ rapidly decreases with $x$ up to $x_c$, and it saturates
when $x$ exceeds $x_c$. Therefore, it is naturally assumed that
$\kappa_s$ is close to zero in the heavily doped samples with $x$
$>$ $x_c$. This result suggests some correlation between the
mobility of the spin excitations and the SP ordering.

Following the above interpretation, it is possible to estimate
$\kappa_s$ by subtracting $\kappa$ of the $x = 0.040$ sample from
the measured $\kappa$ for each sample. The result indicates that
most part of the $\kappa$ of pure CuGeO$_3$ is due to $\kappa_s$,
which is approximately 0.4~W/cm K at 17.5~K. This gives the
diffusivity of the spin excitations $D_s$ of
$\sim$1.6$\times$10$^{-3}$~m$^2$/s, if we assume $C_s$
$\sim$1.8~J/mol K.\cite{ando} $D_s$ in 1D systems can be written
as $v_s l_s$, where $v_s$ is the velocity of the spin excitations
and $l_s$ is the mean free path, in other words, the spin
diffusion length. Assuming the spin-wave like dispersion given by
des Cloizeaux and Pearson \cite{cloizeaux} $\epsilon$=($\pi
/2$)$J_c |\sin (kc)|$ and $J_c$ $\sim$ 120~K, \cite{nishi} one can
evaluate $v_s$ $\sim$ ($\pi /2$)$c J_c$/$\hbar$=
1.3$\times$10$^{4}$~m/s and $l_s$ $\sim$ 130~nm ($\epsilon$ is the
energy of the spin excitations, $k$ is the wave vector and $c$
($\sim$ 0.3~nm) is the distance between adjacent spins). Note that
the spin diffusion length of the same order of magnitude was
obtained from the NMR relaxation measurement for another 1D spin
system, AgVP$_2$S$_6$.\cite{takigawa}

Since $l_s/c \sim$~430 is much larger than unity,
we can conclude that the spin heat transport
for the pure CuGeO$_3$ is coherent.
If the same estimation is applied also to the Mg-doped samples,
$l_s/c \sim$ 100 for the $x = 0.016$ ($< x_c$) and
$l_s/c \sim$ 50 even for $x = x_c$ at this temperature.
Since $c/x_c$ is approximately calculated as 40,
the above estimation indicates that
the spin diffusion length in the 1D-QSL state
exceeds the mean impurity distance $c/x$,
as long as $x$ is less than $x_c$.
Whereas, the spin excitations are not so mobile when $x > x_c$.

\subsection{$\kappa$ just above $T_{\rm SP}^\chi$:
short-range SP order}

Figure~5 shows
$\kappa - \kappa(x=0.040)$ ($\equiv$ $\Delta \kappa$)
of the three Mg-doped samples which have SP transition.
One can find that
$\Delta \kappa$ is almost temperature independent above
$T^*$ $\sim$ 15~K and
that $\Delta \kappa$ of all the samples
deviates from this temperature-independent value
below $T^*$.
Since $T^*$ is close to $T_{\rm SP}$
of the pure CuGeO$_3$,
the deviation can be
attributed to a precursor of the SP transition.
Note that similar behavior is observed in the
Raman scattering measurement on Zn- and Si-dopd CuGeO$_3$.
\cite{kuroe2}

\begin{figure}
\includegraphics[width=0.8\columnwidth]{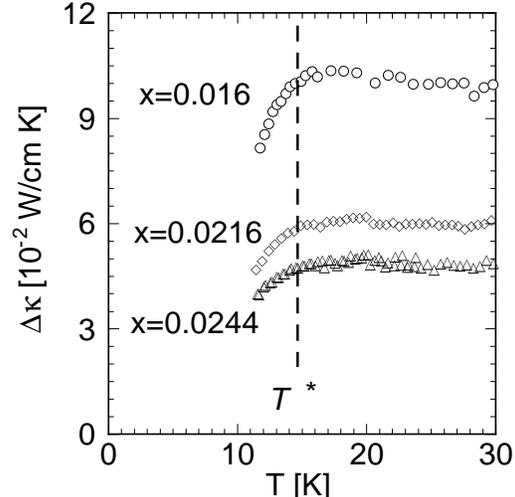}
\caption{Temperature dependence of $\Delta \kappa$ [$\equiv \kappa
- \kappa(x = 0.040)$], for the $x = 0.016$, 0.0216 and 0.0244
samples. The local spin gap opens below $\sim$~15~K in all the
samples. } \label{fig5}
\end{figure}

Recently, it is shown by synchrotron x-ray diffraction
measurement that
FWHM of the Bragg peak from the lattice
dimerization reaches the resolution limit at a temperature
$T_{\rm SP}^{\rm x-ray}$ which is far below $T_{\rm SP}^\chi$.
\cite{wang}
The authors claimed that
the long-range order takes place only below
$T_{\rm SP}^{\rm x-ray}$ and that
lattice dimerization is short-range
just below $T_{\rm SP}^\chi$.\cite{wang}
The gap-like feature in $\kappa$ is also to be explained
along the idea of the short-range order (SRO).
However, the data in Fig.~5 suggests that
SRO grows from $T^*$ $\sim$ 15~K,
which is even above $T_{\rm SP}^{\rm \chi}$.

The field dependence of $\kappa$ is a strong indication
of the presence of the SRO below $T^*$.
Figure~6 shows the temperature dependence of $\Delta \kappa$
for Mg-doped samples in various magnetic fields.
In all cases, we observed the recovery of $\Delta \kappa$
with increasing fields below $T^*$.
The results are consistent with the notion that the reduction of
$\Delta \kappa$ below $T^*$ is due to the development
of the local spin gap.
Since the magnetic field is thought to reduce the magnitude of
spin gap even above $T_{\rm SP}^\chi$,
the number of spin excitations responsible for the heat transport
will increase with the field.
Moreover, the field dependence only appears below $T^*$
in our observation.
We have no field dependence above $T^*$,
which is consistent with the notion
that even the local spin gap is absent.

\begin{figure}
\includegraphics[width=0.8\columnwidth]{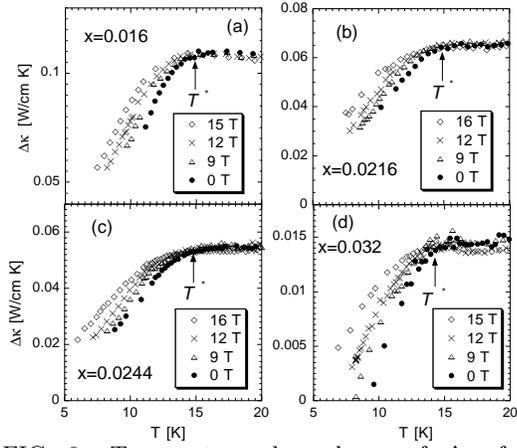}
\caption{Temperature dependence of $\Delta \kappa$ for the (a) $x
= 0.016$, (b) $x = 0.0216$, (c) $x = 0.0244$ and (d) $x = 0.032$
samples. Magnetic fields are applied along the chain direction.}
\label{fig6}
\end{figure}

It should be emphasized that the heat transport by spin
excitations is very sensitive to the spin-Peierls SRO. We observed
a magnetic field dependence in $\Delta \kappa$ below $T^*$ even
for the sample with $x = 0.032\ (> x_c)$, where the spin-Peierls
LRO no longer exist in the whole temperature range. According to
the synchrotron x-ray measurement, SP-SRO develops below 7.5~K in
the $x = 0.032$ sample. \cite{wang} For the $x = 0.040$ sample,
however, there was no evidence that SP-SRO actually develops. In
contrast, one can see that the magnetic field dependence of
$\kappa$ still exists for the $x = 0.040$ sample (Fig.~1(f)) even
though the change is not so obvious as in the $x = 0.032$ sample.
We think that the difference of a length scale of probed area
between the two measurement techniques causes such discrepancy.

The above results tell us that
the impurity-substitution modifies
the ground state and suppresses the spin-Peierls
ordering in CuGeO$_3$ system
through different mechanisms.
The importance of the inter-chain coupling
and the tendency to the 3D-antiferromagnetism increases
below $\sim T_N$.
As a result,
$T_N$ increases with $x$ and
the ground state changes from D-AF to U-AF at $x_c$.
\cite{saito1}
On the other hand,
the local spin-gap formation, which occurs at $T^*$,
is goverened only by
one-dimensional nature of the spin system
coupled with 3D phonons and
has nothing to do with 3D-AF fluctuation.
Therefore,
the temperature of SP-SRO is not modified with $x$,
as long as the spin-singlet state
is energetically favored.
(the spin-singlet state will be no more favored in a
heavily disordered system where long-wave-length
spin excitations, whose energy is less than the spin-gap energy,
are absent.\cite{uhrig})
The spin-diffusion length $l_s$, in turn, rapidly diminishes with $x$
owing to the growing impurity-scattering of the spin excitations.
Since the spin correlation length is directly related to
$l_s$,\cite{halperin}
the length scale of the SP domains
is reduced with $x$.
As a result,
signals of the spin gap,
detected with any probe,
diminishes in size with $x$.

\section{Summary}

The thermal conductivity of the Mg-doped CuGeO$_3$
helps to understand both the spin-gapped state
below the SP transiton and the 1D-QSL state above the
transition.
Large spin heat transport is observed
in the 1D-QSL state for pure CuGeO$_3$,
and rapidly diminishes with Mg-doping,
accompanied by the suppression of the long-range SP
ordering.
The spin gap opening suppresses $\kappa_s$ and
enhances $\kappa_{\rm ph}$.
Examining the $x$ dependence of the spin-gap features,
it turned out that the local spin gap opens at a temperature ($T^*$)
independent of $x$,
suggesting that the suppression of SP ordering is
attributed to the reduction of the spin diffusion
due to the impurity scattering.
It is expected that the above analysis of
the impurity-substitution effect on the thermal conductivity
is applicable to other one-dimensional spin systems
and that thus obtained transport properties may
reveal new aspects on such materials.

\section{Acknowledgment}
We thank A. Kapitulnik for helpful advices both on the experimental
techniques and in understanding the results
in the early stage of this project.

\appendix
\section*{Phonon scattering rate due to point defects}

We can show that the scattering rate $1/\tau_p$ by point defects,
which is introduced by the substitution with the ion with different
mass,
is very small even for the most heavily doped sample
($x = 0.040$).
The dominant-phonon approximation gives,\cite{berman}

\begin{equation}
1/\tau_p \sim \frac{n a^3 (k_B T)^4}{4 m \pi v^3 \hbar ^4 (\Delta M /
M)^2}.
\end{equation}

\noindent
[$n$ ($= x = 0.040$) is density ratio of point defects,
%
$a$ ($\sim 4 \times 10^{-8}$ cm) is lattice constant,
%
$m$ ($=15$) is number of phonon modes
(3 times the number of atoms per unit
cell),
%
$v$ ($\sim 5 \times 10^5$ cm/s) is phonon velocity and
%
$\Delta M/M$ ($= 0.62$) is the ratio
(mass difference between Mg and Cu atoms)/(mass of Cu atom)].
%
$1/\tau_p$ can be estimated to be about $1.0 \times 10^8$
(s$^{-1}$) at 30 K.

In comparison, the total scattering rate $1/\tau$ for
the $x = 0.040$ sample can be obtained from the data as

\begin{equation}
1/\tau \sim \frac{C_{\rm ph} v^2}{3 \kappa_{\rm ph}},
\end{equation}

\noindent
assuming phonon specific heat $C_{\rm ph}$ to be $\sim \beta T^3$ and using
the value of $\beta \sim 2.8 \times 10^{-6}$ (J/cm$^3$ K$^4$),
shown in Ref.~\onlinecite{ando}.
Taking $\kappa_{\rm ph} \sim 0.1$ (W/cm K) from our
result of the $\kappa$ measurement,
$1/\tau \sim 6.4 \times 10^{10}$ (s$^{-1}$) at 30 K.
Since this $1/\tau$ is more than 600 times
larger than $1/\tau_p$ (the ratio is even larger below 30 K),
the scattering by the point defects
is not important for the three-dimensional phonon heat transport
in the measured temperature range.

\end{multicols}


\begin{references}
\bibitem{defect}
It is hardly possible to assume that
defect scattering is mainly responsible in $\kappa_{\rm ph}$
in the 1D-QSL state and that it causes
the parallel shift in the
$\kappa(T)$ curves for the $x \geq 0.016$ samples,
because such scattering provides $\kappa_{\rm ph}(T)$
proportional to $T^2$.
\end{references}
\end{document}